\documentclass[12pt]{article}
\usepackage{amsmath}

\input tcilatex

\begin{document}

\author{I. Radinschi$^{\text{1}}$ and I-Ching Yang$^{\text{2}}$ \\
%EndAName
$^{\text{1}}$Department of Physics \\
``Gh. Asachi'' Technical University, \\
Iasi, 700050, Romania \\
radinschi@yahoo.com and\\
$^{\text{2}}$Department of Natural Science Education, \\
National Taitung University, Taitung, Taiwan 950, \\
Republic of China, icyang@dirac.phys.ncku.edu.tw, icyang@nttu.edu.tw}
\title{Landau-Lifshitz and Weinberg Energy-Momentum Complexes for $2+1$
Dimensional Black Hole Solutions}
\date{}
\maketitle

\begin{abstract}
The aim of this paper is to evaluate the energy distribution of some $2+1$
black hole solutions applying the Landau-Lifshitz and Weinberg definitions.
The metrics under consideration describe the charged black hole, the
solution coupling to a static scalar field and the static and circularly
symmetric exact solution of the Einstein-massless scalar equation. Further,
we compare the expressions for energy with those obtained using the Einstein
and M\o ller prescriptions and give a discussion of the results.


PACS: 04. 20.-q, 04. 20.
\end{abstract}

\section{Introduction}

The subject of energy-momentum localization\ lacks of a definite answer and
continues to be one of the most interesting and challenging problem of
general relativity. Since Einstein [1] has given his important expression
for the energy-momentum complex [1], the localization of energy is also
connected to the use of other energy-momentum complexes, including the
prescriptions of Landau-Lifshitz [2], Papapetrou [3], Bergmann-Thomson [4],
Weinberg and Qadir-Sharif [5] and M\o ller [6]. These prescriptions have
been criticized because of their drawback, they are coordinate dependent.
Except the M\o ller energy-momentum complex which enables one to calculate
the energy distribution in any coordinate system, the other prescriptions
give meaningful results only if the calculations are restricted to
quasi-Cartesian coordinates.

In the recent years the issue of the energy-momentum localization by using
the energy-momentum complexes was re-opened. Many researchers studied
different space-times and obtained interesting results, which demonstrate
that these definitions are powerful concepts for energy-momentum
localization. In 1990 Bondi [7] gave his opinion that "a nonlocalizable form
of energy is not admissible in general relativity, because any form of
energy contributes to gravitation and so its location can in principle be
found". Misner et al [8] sustained that to look for a local energy-momentum
means that is looking for the right answer to the wrong question. They also
concluded that the energy is localizable only for spherical systems. On the
other hand, Cooperstock and Sarracino [9] demonstrated that if the energy is
localizable in spherical systems then it is also localizable in any
space-times. In this context, of great importance is the Cooperstock
hypothesis [10] which states that energy and momentum are confined to the
regions of non-vanishing energy-momentum tensor for the matter and all
non-gravitational fields.

The problem of energy-momentum localization by applying the energy-momentum
complexes was revived at the beginning of the last decade and many
interesting results have been obtained [11]. Virbhadra [11] showed that the
Einstein, Landau-Lifshitz, Papapetrou and Weinberg energy-momentum complexes
give the same energy density as the Penrose definition for a general
non-static spherically symmetric metric of the Kerr-Schild class.
Furthermore, important works were done with the energy-momentum complexes in
2- and 3-dimensional space-times [12]. Considerable investigations have been
performed in elaborating interesting papers, which demonstrate that
energy-momentum complexes yield the same results as their tele-parallel
versions for a given space-time [13]. All these considerations point out the
significance of these prescriptions and stress the usefulness of
energy-momentum complexes for energy-momentum localization.

In our paper we evaluate the energy distribution of three $2+1$ dimensional
black hole solutions applying the Landau-Lifshitz and Weinberg\ definitions.
The paper is organized as follows. In Sec. 2 we briefly present the $2+1$
dimensional black hole solutions that we study. In Sec. 3 we give a
description of the Landau-Lifshitz and Weinberg energy-momentum complexes
and we compute the energy distributions for the three $2+1$ dimensional
black hole solutions, and also make a comparison with the values of energy
obtained in the Einstein and M\o ller prescriptions. Finally, in Sec. 4 we
make a summary of the obtained results and some concluding remarks. The
metrics under consideration describe the charged black hole [14], the
solution coupling to a static scalar field [14] and the static and
circularly symmetric exact solution of the Einstein-massless scalar equation
[15]. Through the paper we follow the convention that Latin indices run from 
$0$ to $2$ and the Greek indices run from $1$ to $2$.

\section{Black Hole Solutions in $2+1$ Dimensions}

In recent years the Einstein theory of gravity in $2+1$ dimensions has
gained considerable attention [16]. Due to an expectation is that the study
of $2+1$ dimensional theories would provide relevant information about the
corresponding theory in $3+1$ dimensions. In $2+1$ dimensions the number of
independent components of the Riemann curvature tensor and the Einstein
tensor are the same, consequently the imposition of Einstein's equations in
vacuum implies that the curvature tensor also vanishes and there are no
gravitational waves and no interactions between masses. Therefore, the
space-time described by the vacuum solutions to Einstein's equations in $2+1$
dimensions allows that the existence of the black holes would be prevented
[17]. However, Ba\~{n}ados, Teitelboim and Zanelli (BTZ) [18] have
discovered a black hole solution to the EM equations (with a negative
cosmological constant) in $2+1$ dimensions, which is characterized by mass,
angular momentum and charge parameters.

Because the energy-momentum complexes are fundamental quantities used for
energy-momentum localization, we study the energy distributions of some
black hole solutions in $2+1$ dimensions applying the Landau-Lifshitz and
Weinberg prescriptions.

The metrics under consideration in our paper are:

i) Charged black hole [14], which is expressed by the line element

\begin{equation}
ds^{2}=-(\Lambda \,r^{2}-M\,-2\,Q^{2}\,\ln (\frac{r}{r_{+}}))dt^{2}+(\Lambda
\,r^{2}-M\,-2\,Q^{2}\,\ln (\frac{r}{r_{+}}))^{-1}\,dr^{2}+r^{2}\,d\varphi
^{2},  \tag{1}
\end{equation}

where $r_{+}=\sqrt{\frac{M}{\Lambda }}$.

ii) Solution coupling to a static scalar field [14] which is described by
the metric

\begin{equation}
ds^{2}=-(\frac{(r-2\,B)(B+r)^{2}\,\Lambda }{r})dt^{2}+(\frac{%
(r-2\,B)(B+r)^{2}\,\Lambda }{r})^{-1}dr^{2}+r^{2}\,d\varphi ^{2},  \tag{2}
\end{equation}

where $B$ is a constant and the scalar field is given by $\Phi =\sqrt{\frac{B%
}{\pi (r+B)}}$.

iii) The static and circularly symmetric exact solution of the
Einstein-massless scalar equation [15] which is given by

\begin{equation}
ds^{2}=-B\,dt^{2}+B^{-1}\,dr^{2}+r^{2}\,d\varphi ^{2},  \tag{3}
\end{equation}

where $B=(1-q)R^{q}$, $R=\frac{r}{r_{0}}$, $q$ is the scalar charge and the
scalar field is given by $\Phi =\sqrt{\frac{q}{k}}\ln R$. The case $q=0$
corresponds to the flat space-time in $2+1$ dimensions.

For these three $2+1$ dimensional black hole solutions we compute the energy
distributions using the Landau-Lifshitz and Weinberg\ prescriptions.

\section{Landau-Lifshitz and Weinberg Prescriptions for $2+1$ Dimensional
Black Hole Solutions}

The Landau-Lifshitz energy-momentum complex [2] is given by

\begin{equation}
L^{i\,k}=\frac{1}{2\,k}S^{i\,k\,l\,m},_{l\,m},  \tag{4}
\end{equation}

where $k$ is the coupling gravitational constant and

\begin{equation}
S^{i\,k\,l\,m}=-g(g^{i\,k}\,g^{l\,m}-g^{i\,l}\,g^{k\,m}).  \tag{5}
\end{equation}

$L^{0\,0}$ and $L^{\alpha \,0}$ are the energy and momentum density
components, respectively.

The Landau-Lifshitz energy-momentum complex satisfies the local conservation
law

\begin{equation}
\frac{\partial L^{i\,k}}{\partial x^{k}}=0.  \tag{6}
\end{equation}

Using Gauss's theorem, the energy and momentum components are

\begin{equation}
P^{i}=\frac{1}{2\,k}\diint S^{i\,0\,\alpha \,m},_{m}\,n_{\alpha }\,dS, 
\tag{7}
\end{equation}

where $n_{\alpha }=\left( {\frac{x}{r}},{\frac{y}{r}},{\frac{z}{r}}\right) $
are the components of a normal vector over an infinitesimal surface element.

The energy and momentum for a three dimensional background are given by

\begin{equation}
P^{i}=\diint L^{i\,0}\,dx^{1}\,dx^{2}.  \tag{8}
\end{equation}

The Weinberg energy-momentum complex [5] is given by

\begin{equation}
W^{i\,k}=\frac{1}{2\,k}D_{\;\,\;\;\,,l}^{l\,i\,k},  \tag{9}
\end{equation}

where

\begin{equation}
D_{\;\,\;}^{l\,i\,k}=\frac{\partial h_{a}^{a}}{\partial x_{l}}\eta ^{i\,k}-%
\frac{\partial h_{a}^{a}}{\partial x_{i}}\eta ^{l\,k}-\frac{\partial h^{a\,l}%
}{\partial x^{a}}\eta ^{i\,k}+\frac{\partial h^{a\,i}}{\partial x^{a}}\eta
^{l\,k}+\frac{\partial h^{l\,k}}{\partial x_{i}}-\frac{\partial h^{i\,k}}{%
\partial x_{l}},  \tag{10}
\end{equation}

with

\begin{equation}
h_{i\,k}=g_{i\,k}-\eta _{i\,k},  \tag{11}
\end{equation}

where $\eta ^{i\,k}=diag(-1,1,1)$ and $W^{0\,0}$ and $W^{\alpha \,0}$ are
the energy and the momentum density components, respectively.

The Weinberg energy-momentum complex satisfies the local conservation law

\begin{equation}
\frac{\partial W^{i\,k}}{\partial x^{k}}=0.  \tag{12}
\end{equation}

Applying Gauss's theorem, the energy and momentum components are

\begin{equation}
P^{i}=\frac{1}{2\,k}\diint D^{i\,0\,\alpha }\,n_{\alpha }\,dS,  \tag{13}
\end{equation}

where $n_{\alpha }=\left( {\frac{x}{r}},{\frac{y}{r}},{\frac{z}{r}}\right) $
are the components of a normal vector over an infinitesimal surface element.

The energy and momentum in the Weinberg prescription for a three dimensional
background are given by

\begin{equation}
P^{i}=\diint W^{i\,0}\,dx^{1}\,dx^{2}.  \tag{14}
\end{equation}

For carrying out the calculations with the Landau-Lifshitz and Weinberg\
energy-momentum complexes we transform the general metric given by

\begin{equation}
ds^{2}=-v(r)\,dt^{2}+w(r)\,dr^{2}+r^{2}\,d\varphi ^{2},  \tag{15}
\end{equation}

to quasi-Cartesian coordinates $t,x,y$\ by using $x=r\,\cos \varphi $, $%
y=r\,\sin \varphi $ and we obtain

\begin{equation}
ds^{2}=-v\,dt^{2}+(dx^{2}+dy^{2})+\frac{w-1}{x^{2}+y^{2}}(xdx+ydy)^{2}. 
\tag{16}
\end{equation}

The determinant of the metric (16) is $g=-v\,w$ and the covariant components
of the metric tensor are given by

\begin{equation}
g_{i\,k}=%
\begin{vmatrix}
\frac{x^{2}\,w+y^{2}}{x^{2}+y^{2}} & \frac{x\,y(w-1)}{x^{2}+y^{2}} & 0 \\ 
\frac{x\,y(w-1)}{x^{2}+y^{2}} & \frac{y^{2}\,w+x^{2}}{x^{2}+y^{2}} & 0 \\ 
0 & 0 & -v%
\end{vmatrix}%
.  \tag{17}
\end{equation}

For the contravariant components of the metric tensor we obtain

\begin{equation}
g^{i\,k}=%
\begin{vmatrix}
\frac{y^{2}\,w+x^{2}}{(x^{2}+y^{2})w} & -\frac{x\,y(w-1)}{(x^{2}+y^{2})w} & 0
\\ 
-\frac{x\,y(w-1)}{(x^{2}+y^{2})w} & \frac{x^{2}\,w+y^{2}}{(x^{2}+y^{2})w} & 0
\\ 
0 & 0 & -\frac{1}{v}%
\end{vmatrix}%
.  \tag{18}
\end{equation}

The required nonvanishing components $S^{i\,0\,\alpha }$ of the
Landau-Lifshitz energy-momentum complex are

\begin{equation}
S^{0\,0\,1}=\frac{x(1-w)}{x^{2}+y^{2}},S^{0\,0\,2}=\frac{y(1-w)}{x^{2}+y^{2}}%
\,.  \tag{19}
\end{equation}

For the Weinberg prescription, the required nonvanishing components $%
D^{i\,0\,\alpha }$ are given by%
\begin{equation}
D^{0\,0\,1}=\frac{x(1-w)}{x^{2}+y^{2}},\,D^{0\,0\,2}=\frac{y(1-w)}{%
x^{2}+y^{2}}.  \tag{20}
\end{equation}

Using (8), (14), (18), (19), (20), applying Gauss's theorem and after
performing the calculations we obtain that the energy within a circle with
radius $r$ in the Landau-Lifshitz and Weinberg prescriptions is given by

\begin{equation}
E_{LL}=E_{W}=\frac{1}{2\,k}\doint (1-w)d\varphi .  \tag{21}
\end{equation}

We obtain that the expression of energy in the Landau-Lifshitz prescription
is the same as in the Weinberg prescription.

For the aforementioned black hole solutions in $2+1$ dimensions we obtain
the next results in the Landau-Lifshitz and Weinberg prescriptions:

i) For the charged black hole $v=\Lambda \,r^{2}-M\,-2\,Q^{2}\,\ln (\frac{r}{%
r_{+}})$, $w=(\Lambda \,r^{2}-M\,-2\,Q^{2}\,\ln (\frac{r}{r_{+}}))^{-1}$ and
the energy distribution computed with the Landau-Lifshitz and Weinberg
prescriptions is given by

\begin{equation}
E_{LL}=E_{W}=\frac{\pi }{k}(\frac{\Lambda \,r^{2}-M-2\,Q^{2}\,\ln (\frac{r}{%
r_{+}})-1}{\Lambda \,r^{2}-M-2\,Q^{2}\,\ln (\frac{r}{r_{+}})}).  \tag{22}
\end{equation}

ii) In the case of the solution coupling to a static scalar field $v=\frac{%
(r-2\,B)(B+r)^{2}\,\Lambda }{r}$, $w=(\frac{(r-2\,B)(B+r)^{2}\,\Lambda }{r}%
)^{-1}$ and for the energy we obtain

\begin{equation}
E_{LL}=E_{W}=\frac{\pi \,[\Lambda (r-2\,B)(B+r)^{2}-r]}{k\,\Lambda
(r-2\,B)(B+r)^{2}}.  \tag{23}
\end{equation}

iii) For the static and circularly symmetric exact solution of the
Einstein-massless scalar equation $v=(1-q)R^{q}$, $w=((1-q)R^{q})^{-1}$ and
the Landau-Lifshitz and Weinberg energy-momentum complexes yield for the
energy distribution the expression

\begin{equation}
E_{LL}=E_{W}=\frac{\pi }{k}[\frac{(1-q)R^{q}-1}{(1-q)R^{q}}].  \tag{24}
\end{equation}

Some remarks are needed. In a previous work [19] we computed the energy
distributions of these three $2+1$ dimensional black hole solutions applying
the Einstein and M\o ller prescriptions and we obtained

\begin{equation}
E_{E}=\frac{1}{2\,k}\doint \frac{v}{\sqrt{v\,w}}(1-w)d\varphi  \tag{25}
\end{equation}

and

\begin{equation}
E_{M}=-\frac{1}{k}\doint \frac{r}{\sqrt{v\,w}}\frac{\partial v}{\partial r}%
d\varphi ,  \tag{26}
\end{equation}

respectively.

We make a comparison with the values of energy obtained using the
Landau-Lifshitz and Weinberg energy-momentum complexes. In the case of the
all three $2+1$ dimensional black hole solutions we conclude that between
the Einstein, Landau-Lifshitz and Weinberg prescriptions there is a
relationship given by

\begin{equation}
E_{LL}=E_{W}=w\,E_{E}.  \tag{27}
\end{equation}

M\o ller's energy-momentum complex yields different results for the energy
distribution of the aforementioned $2+1$ dimensional black hole solutions
than the Einstein, Landau-Lifshitz and Weinberg prescriptions.

It is important that the expression for the energy obtained in the
Landau-Lifshitz prescription exactly matches with that computed applying the
Weinberg prescription. Even these definitions of Einstein, Landau-Lifshitz,
Weinberg and M\o ller\ do not provide the same result for the energy
distribution, we point out that the connections between the expressions for
energy obtained in the Einstein, Landau-Lifshitz and Weinberg prescriptions
are similar to\ the $3+1$ dimensional case [20] (see therein eqs. 39-42, for 
$A=B^{-1}$, $D=1$ and $F=0$), when the calculations are done in
Schwarzschild-Cartesian coordinates.

\section{Discussion}

For solving the problem of energy and momentum localization, many attempts
have been made in the past but this remains an important issue to be
settled. The difficulty relies in the lack of a generally accepted
expression for the energy density. Even the energy-momentum complexes
\textquotedblright seem\textquotedblright\ to be useful for the localization
of energy, there are doubts that these prescriptions could give acceptable
results for a given space-time. The results obtained by several authors
[11]-[13] demonstrated that the energy-momentum complexes are good tools for
evaluating the energy and momentum in general relativity, and working with
them we can obtain acceptable expressions for the energy associated with a
given space-time. Chang, Nester and Chen [21] showed that the
energy-momentum complexes are actually quasi-local and legitimate expression
for the energy-momentum. They concluded that there exist a direct
relationship between energy-momentum complexes and quasi-local expressions
because every energy-momentum complexes is associated with a legitimate
Hamiltonian boundary term. Their idea supports the energy-momentum complexes
and the role which these are playing in energy-momentum localization.
Furthermore, important studies have been done about the new idea of
quasi-local approach for energy-momentum complexes [21]-[22] and a large
class of new pseudotensors connected to the positivity in small regions have
been studied and constructed [23]. In this light, the quasi-local quantities
are associated with a closed 2-surface (L. B. Szabados, [22] and
http://relativity.livingreviews.org/Articles/lrr-2004-4/). The Hamiltonian
boundary term determines the quasi-local quantities for finite regions and
the special quasi-local energy-momentum boundary term expressions correspond
each of them to a physically distinct and geometrically clear boundary
condition [24].

In this paper we continue the investigations concerning the energy of some $%
2+1$ dimensional black hole solutions. We evaluate the energy distribution
for three $2+1$ dimensional black hole solutions using the Landau-Lifshitz
and Weinberg\ prescriptions. Furthermore, we compare our result with those
obtained in the Einstein and M\o ller prescriptions and investigate the
connections between the expressions for the energy obtained with these
energy-momentum complexes. The Landau-Lifshitz and Weinberg prescriptions
yield the same expressions for the energy distribution of the aforementioned 
$2+1$ dimensional black hole solutions, sustaining the viewpoint that
different energy-momentum complexes can give the same result for a given
space-time. The connection between the expressions for the energy
distribution obtained in these three prescriptions, Einstein,
Landau-Lifshitz and Weinberg is given by the relationship $%
E_{LL}=E_{W}=wE_{E}$. It is important to notice that the connections between
the expressions for energy obtained in the Einstein, Landau-Lifshitz and
Weinberg prescriptions are similar to the $3+1$ dimensional case [20] (see
therein eqs. 39-42, for $A=B^{-1}$,$\ D=1$ and $F=0$), when the calculations
are done in Schwarzschild-Cartesian coordinates.

As $r$ becomes larger, the energy distributions of the three $2+1$
aforementioned dimensional black hole solutions become finite. Furthermore,
in the case $r\rightarrow \infty $ the energy distributions of these
solutions do not diverge. These energy distributions do not diverge because
Einstein's and M\o ller's energy complexes are covariant, but
Landau-Lifshitz and Weinberg's energy complexes are contravariant. These
three black hole solutions are not asymptotically flat, so the covariant
energy complex will be divergent. However, the contravariant energy complex
will be not divergent.

Our paper extends a previous study [19] about the energy of $2+1$
dimensional black hole solutions and sustains the viewpoint that the
energy-momentum complexes are important concepts for energy-momentum
localization. Furthermore, our work also supports a) the opinion that
different energy-momentum complexes can yield the same expression for the
energy in a given space-time and b) the connection between the values of
energy obtained applying the Einstein, Landau-Lifshitz and Weinberg
prescriptions are similar to the $3+1$ dimensions when the calculations are
done in Schwarzschild-Cartesian coordinates. An open question remains, why
these prescriptions (ELLPW) and M\o ller do not allow obtaining the same
expression for the energy distribution. We conclude that we obtained
different results applying the definitions of Einstein, Landau-Lifshitz and
Weinberg because these energy-momentum complexes are pseudotensors and are
non-covariant, coordinate dependent expressions [21]-[24] and this agrees
with the equivalence principle which states that gravity cannot be detected
at a point.


\begin{thebibliography}{99}
\bibitem{1} A. Einstein, \textit{Preuss. Akad. Wiss. Berlin }\textbf{47},
778 (1915); Addendum-ibid. \textbf{47}, 799 (1915); A. Trautman, in \textit{%
Gravitation: an Introduction to Current Research}, ed. L. Witten (Wiley, New
York, 1962, p. 169).

\bibitem{2} L. D. Landau and E. M. Lifshitz, \textit{The Classical Theory of
Fields }(Pergamon Press, 1987, p. 280).

\bibitem{3} A. Papapetrou, \textit{Proc. R. Irish. Acad.} \textbf{A52}, 11
(1948).

\bibitem{4} P. G. Bergmann and R. Thomson, \textit{Phys. Rev.} \textbf{89},
400 (1953).

\bibitem{5} S. Weinberg, \textit{Gravitation and Cosmology: Principles and
Applications of General Theory of Relativity} (John Wiley and Sons, Inc.,
New York, 1972, p. 165); A. Qadir and M. Sharif, \textit{Phys. Lett.} 
\textbf{A167}, 331 (1992).

\bibitem{6} C. M\o ller, \textit{Ann. Phys. (NY)} \textbf{4}, 347 (1958).

\bibitem{7} H. Bondi, \textit{Proc. R. Soc. London} \textbf{A427}, 249
(1990).

\bibitem{8} C. W. Misner, K. S. Thorne and J. A. Wheeler, Gravitation, W. H.
Freeman and Co., NY , 603 (1973).

\bibitem{9} F. I. Cooperstock and R. S. Sarracino, \textit{J. Phys. A: Math.
Gen.} \textbf{11}, 877 (1978).

\bibitem{10} F. I. Cooperstock, \textit{Mod. Phys. Lett. }\textbf{A14}, 1531
(1999).

\bibitem{11} K. S. Virbhadra, \textit{Phys. Rev.} \textbf{D41}, 1086 (1990);
K. S. Virbhadra, \textit{Phys. Rev.} \textbf{D42}, 2919 (1990); N. Rosen and
K. S. Virbhadra, \textit{Gen. Rel. Grav.} \textbf{25}, 429 (1993); K. S.
Virbhadra and J. C. Parikh, \textit{Phys. Lett.} \textbf{B331}, 302 (1994);
A. Chamorro and K. S. Virbhadra, \textit{Pramana - J. Phys.} \textbf{45},
181 (1995); J. M. Aguirregabiria, A. Chamorro and K. S. Virbhadra, \textit{%
Gen. Rel. Grav.} \textbf{28}, 1393 (1996); A. Chamorro and K. S. Virbhadra, 
\textit{Int. J. Mod. Phys. }\textbf{D5}, 251 (1997); K. S. Virbhadra, 
\textit{Phys. Rev.} \textbf{D60}, 104041 (1999); S. S. Xulu, \textit{Int. J.
Theor. Phys.} \textbf{37}, 1773 (1998); S. S. Xulu, \textit{Int. J. Mod.
Phys. }\textbf{D7}, 773 (1998); E. C. Vagenas, \textit{Int. J. Mod. Phys.} 
\textbf{A18}, 5949 (2003) ; E. C. Vagenas, \textit{Mod. Phys. Lett.} \textbf{%
A21}, 1947 (2006); E. C. Vagenas, \textit{Mod. Phys. Lett.} \textbf{A19},
213 (2004); I. Radinschi, \textit{Acta Physica Slovaca} \textbf{49(5)}, 789
(1999); I. Radinschi, \textit{Mod. Phys. Lett.} \textbf{A15}, Nos. 11\&12,
803 (2000); I-Ching Yang and I. Radinschi, \textit{Chin. J. Phys.} \textbf{41%
}, 326 (2003); I. Radinschi, \textit{Horizons in World Physics}, Vol. 246, 
\textit{Quantum Cosmology Research Trends}, ed. Albert Reimer, Nova Science
Publishers, Inc New York, U.S.A., 185-198, 2005; I. Radinschi and I-Ching
Yang, \textit{New Developments in String Theory Research}, ed. Susan A.
Grece, Nova Science Publishers, Inc New York, U.S.A., 1-17, 2006; T.
Bringley, \textit{Mod. Phys. Lett. }\textbf{A17}, 157 (2002); M. S\'{u}ken%
\'{\i}k and J. Sima, gr-qc/0101026; M. Sharif and Tasnim Fatima, \textit{%
Int. J. Mod. Phys.} \textbf{A20}, 4309 (2005); M. Sharif, \textit{Nuovo Cim.}
\textbf{B19}, 463 (2004). M. Sharif, \textit{Int. J. Mod. Phys. }\textbf{D13}%
, 1019 (2004); M. Sharif and Tasnim Fatima, \textit{Nuovo Cim.} \textbf{B120}%
, 533 (2005); M. Sharif and Tasnim Fatima, \textit{Astrophys. Space Sci.} 
\textbf{302}, 217 (2006); M. Sharif, M. Azam, gr-qc/0612048, accepted for
publication in \textit{Int. J. Mod. Phys. A}; Ragab M. Gad, \textit{Mod.
Phys. Lett.} \textbf{A19}, 1847 (2004); Ragab M. Gad, \textit{Gen. Rel. Grav}%
. \textbf{38}, 417 (2006); Ragab M. Gad, \textit{Astrophys. Space Sci.} 
\textbf{295}, 451 (2005); Ragab M. Gad, \textit{Astrophys. Space Sci.} 
\textbf{293}, 453 (2004); Ragab M. Gad, \textit{Astrophys. Space Sci.} 
\textbf{295}, 459 (2005); Ragab M. Gad, \textit{Astrophys. Space Sci.} 
\textbf{302} 141 (2006); Ragab M. Gad, gr-qc/0603075, accepted for
publication in \textit{Int. J. Theor.Phys.}; O. Patashnick, \textit{Int. J.
Mod. Phys.} \textbf{D14}, 1607 (2005); M. Salti , \textit{Nuovo Cim.} 
\textbf{120B}, 53 (2005); O. Aydogdu, \textit{Fortsch. Phys.} \textbf{54},
246 (2006); M. Salti, \textit{Astrophys. Space Sci.} \textbf{299}, 159
(2005); O. Aydogdu and M. Salti, \textit{Czech. J. Phys.} \textbf{56}, 789
(2006); O. Aydogdu, M. Salti, M. Korunur, Irfan Acikgoz, \textit{Found.
Phys. Lett.} \textbf{19}, 709 (2006); M. Salti, O. Aydogdu and M. Korunur, 
\textit{JHEP} \textbf{0612}, 078 (2006); P. Halpern, \textit{Astrophys.
Space Sci.} \textbf{306}, 279 (2006).

\bibitem{12} K. S. Virbhadra, \textit{Pramana-J. Phys.} \textbf{44}, 317
(1995); E. C. Vagenas, \textit{Int. J. Mod. Phys.} \textbf{A18}, 5949
(2003); E. C. Vagenas, \textit{Int. J. Mod. Phys.} \textbf{A18}, 5781
(2003); E. C. Vagenas, I\textit{nt. J. Mod. Phys.} \textbf{D14}, 573 (2005);
Th. Grammenos, \textit{Mod. Phys. Lett.} \textbf{A20}, 1741 (2005); I.
Radinschi and Th. Grammenos, \textit{Int. J. Mod. Phys.} \textbf{A21}, 2853
(2006); I-Ching Yang and I. Radinschi, gr-qc/0309130, to appear in AIP

\bibitem{13} Gamal G. L. Nashed, \textit{Phys. Rev.} \textbf{D66}, 064015
(2002); Gamal G. L. Nashed, \textit{Nuovo Cim.} \textbf{117B}, 521 (2002);
Gamal G. L. Nashed, \textit{Int. J. Mod. Phys.} \textbf{A21}, 3181 (2006);
M. Korunur, A. Havare, M. Salti and O. Aydogdu, gr-qc/0502031; M. Salti and
A. Havare, gr-qc/0502042; M. Salti and A. Havare, gr-qc/0502058; M. Salti
and A. Havare, \textit{Int. J . Mod. Phys.} \textbf{A20}, 2169 (2005); M.
Salti, \textit{Mod. Phys. Lett.} \textbf{A20}, 2175 (2005); M. Salti, 
\textit{Int. J. Mod. Phys.} \textbf{D15}, 695 (2006); O. Aydogdu and M.
Salti, \textit{Astrophys. Space Sci.} \textbf{299}, 227 (2005); O. Aydogdu
and M. Salti, \textit{Astrophys. Space Sci.} \textbf{302}, 61 (2006); M.
Salti, \textit{Acta Phys. Slov.} \textbf{55}, 563 (2005); O. Aydogdu, M.
Salti and M. Korunur, \textit{Acta Phys. Slov.} \textbf{55}, 537 (2005); M.
Salti, gr-qc/0607011, to appear in Astrophys. Space Sci.; M. Salti,
gr-qc/0607116; Murat Korunur, Mustafa Salti, Oktay Aydogdu, Irfan Acikgoz,
gr-qc/0607117; Sezgin Aygun, Melis Aygun, Ismail Tarhan, gr-qc/0607103

\bibitem{14} M. Horta\c{c}su, H. T. \"{O}z\c{c}elik and B. Yapi\c{s}kan, 
\textit{Gen. Rel. Grav.} \textbf{35}, 1209 (2003).

\bibitem{15} K. S. Virbhadra, \textit{Pramana-J. Phys.} \textbf{44}, 317
(1995).

\bibitem{16} S. Deser, R. Jackiw and S. Templeton, \textit{Ann. Phys. (NY)} 
\textbf{140}, 372 (1982); A. Ach\'{u}arro and P. K. Townsend, \textit{Phys.
Lett.} \textbf{B180}, 89 (1986); D. Bak, D. Cangemi and R. Jackiw, \textit{%
Phys. Rev.} \textbf{D49}, 5173 (1994); J. D. Brown, J. Creighton and R.
Mann, \textit{Phys. Rev.} \textbf{D50}, 6394 (1994).

\bibitem{17} S. Giddings, J. Abbot and K. Kuchar, \textit{Gen. Rel. Grav.} 
\textbf{16}, 751 (1984).

\bibitem{18} M. Ba\~{n}ados, C. Teitelboim and J. Zanelli, \textit{Phys.
Rev. Lett.} \textbf{69}, 1849 (1992).

\bibitem{19} I-Ching Yang and I. Radinschi, gr-qc/0309130, to appear in AIP

\bibitem{20} K. S. Virbhadra, \textit{Phys. Rev.} \textbf{D60}, 104041
(1999).

\bibitem{21} Chia-Chen Chang, J. M. Nester and Chiang-Mei Chen, \textit{%
Phys. Rev. Lett.} \textbf{83}, 1897 (1999).

\bibitem{22} Chiang-Mei Chen and J. M. Nester, \textit{Class. Quant. Grav.} 
\textbf{16}, 1279 (1999); Chiang-Mei Chen and J. M. Nester, \textit{Grav.
Cosmol.} \textbf{6}, 257 (2000); Chiang-Mei Chen, James M. Nester and Roh
Suan Tung, \textit{Phys. Lett.} \textbf{A203}, 5 (1995); L. B. Szabados, 
\textit{Living. Rev. Relativity} \textbf{7}, 4 (2004).

\bibitem{23} Lau Loi So, James M. Nester and Hsin Chen, gr-qc/0605150, to
appear in Proceedings of the 7th International Conference on Gravitation and
Astrophysics; Lau Loi So and James M. Nester, gr-qc/0612061; S. Deser, J.S.
Franklin and D. Seminara, \textit{Class. Quant. Grav.} \textbf{16}, 2815
(1999).

\bibitem{24} James M. Nester \textit{Class. Quant. Grav.} \textbf{21}, S261
(2004); Chiang-Mei Chen, James M. Nester and Roh-Suan Tung, \textit{Phys.
Rev.} \textbf{D72}, 104020 (2005).
\end{thebibliography}
\end{document}